\documentclass[12pt]{iopart}

\usepackage{iopams}  
\usepackage[utf8]{inputenc}
\usepackage{amssymb}
\expandafter\let\csname equation*\endcsname\relax
\expandafter\let\csname endequation*\endcsname\relax
\usepackage{amsmath}
\usepackage{multirow}

\usepackage[dvips]{graphicx}
\usepackage[dvips]{color}
\usepackage[bf,footnotesize]{caption}
\usepackage[bf]{subfigure}

\renewcommand{\d}{{\rm d}}

\renewcommand{\O}[2][]{\hat{#2} ^{\vphantom{\dagger}}_{#1}}
\newcommand{\Op}[2][]{\hat{#2} ^{\dagger}_{#1}}

\begin{document}
\newpage

\title[Superfluid clusters, percolation and phase transitions in the disordered BHM]{Superfluid clusters, percolation and phase transitions in the disordered, two dimensional \mbox{Bose-Hubbard model}}

\author{A.\,E.\,Niederle and H.\,Rieger}

\address{Theoretical Physics, Saarlandes University , D--66041 Saarbr\"ucken, Germany}
\ead{astrid@lusi.uni-sb.de, h.rieger@mx.uni-saarland.de}
\begin{abstract}
The Bose glass (BG) phase is the Griffiths region of the disordered Bose Hubbard model (BHM), characterized by finite, quasi-superfluid clusters within a Mott insulating background. We propose to utilize this characterization to identify the complete zero-temperature phase
diagram of the disordered BHM in $d\ge2$ dimensions by analyzing the geometric properties of what we call superfluid (SF) clusters, which are defined to be clusters of sites with non-integer expectation values for the local boson occupation number. The Mott insulator (MI) phase then is the region in the phase diagram where no SF clusters exist, and the SF phase the region, where SF clusters percolate - the BG phase is in between: SF clusters exist, but do not percolate. This definition is particularly useful in the context of local mean field (LMF, or Gutzwiller-Ansatz) calculations, where we show that an identification of the phases on the basis of global quantities like the averaged SF order parameter and the compressibility are misleading. We apply the SF cluster analysis to the LMF ground states of the two dimensional disordered BHM to produce its phase diagram and find a) an excellent agreement with the phase diagram predicted on the basis of quantum Monte Carlo simulations for the 
commensurate density $n=1$, and b) large differences to stochastic mean field and other mean field predictions for fixed disorder strength. The relation of the percolation transition of the SF clusters with the onset of non-vanishing SF stiffness indicating the BG to SF transition is discussed.
\end{abstract}
\maketitle

\section{Introduction}\label{section:Introduction}
The experimental proof of the Mott insulator (MI) to superfluid (SF) transition in ultracold atomic
systems \cite{Grei02} opened a wide field of interesting research in
this field. In particular the influence of disorder on a system of
bosons in a regular (e.g. optical) lattice received much interest
since the fundamental work of Fisher et al. \cite{Fish89}. Here the
phase diagram and transitions for bosons in a disordered potential was
analysed and the existence of a Bose glass (BG) phase was predicted. The BG
represents a non-SF but, in contrast to the MI, compressible
phase displaying an excitation spectrum with arbitrarily small
excitation energies. The BG phase is the analogue of the
Griffith regions occurring for instance in disordered magnets,
whose physics is dominated by rare region effects due to arbitrarily
large strongly coupled clusters \cite{Fish99,Kisk97,Vojt06}. 

Studies of the excitation spectrum of the disordered system in
dependence of the disorder strength and time-of-flight measurements
confirmed the predicted BG phase experimentally \cite{Fall07}. In
addition to the well controllable optical lattice, disorder is
introduced either by a non-commensurate periodic potential
\cite{Fall07} or by speckle potentials \cite{Lye05, Whit09}. A new
view on the properties of ultra cold bosonic gases opened up as
high resolution techniques allowed access to single-site detection
recently \cite{Bakr10, Sher10}. This progress now yields a direct view
on the population numbers within the different phases and the in-situ
hopping dynamics of the bosons in their optical potential.

Theoretically the phase diagram of the disordered Bose Hubbard model (BHM) has been studied by various methods: The strong coupling expansion \cite{Free96} is a perturbative method up to third order in the tunnelling rate yielding a prediction on the Mott lobes. The disordered BHM was widely studied by quantum Monte Carlo (QMC) methods in various dimension \cite{Soey11,Poll09,Prok04,Lee01,Kisk97,Kisk97a,Krau91}. In addition, density-matrix renormalization group (DMRG) techniques were applied to 1D systems containing either quasi-periodic potentials \cite{Roux08, Deng08, Deng09} or a uniform distribution of disorder strength \cite{Carr10,Raps99}. A frequently used alternative approach is the local mean field (LMF) approximation \cite{Shes93}, which replaces the nearest neighbour hopping on the lattice by isolated bosonic degrees of freedom interacting via an effective mean field coupling with the neighbours. Based on the LMF approximation several numerical techniques, such as stochastic mean field (SMF) 
theory \cite{Biss09, Biss10} and LMF theory \cite{Buon09,Buon07,Buon04b}, were proposed.

An intriguing question has been for a long time the potential existence of a direct MI-SF transition \cite{Soey11,Poll09,Prok04,Lee01,Kisk97,Kisk97a,Krau91}, which is now excluded by a rigorous theorem \cite{Poll09}. The occurrence of the BG phase intervening between the MI and SF is caused by Griffiths effects \cite{Grif69,Fish89} due to arbitrarily large, but exponentially rare clusters of one phase within a background of another phase \cite{Fish95,Rieg96,Fish99}. Since any exponentially rare event is hard to sample numerically, the existence of an intervening BG phase might have been eluded some studies, be it QMC \cite{Prok04,Lee01,Kisk97,Kisk97a,Krau91}, LMF \cite{Buon09,Buon07,Buon04b} or SMF theory \cite{Biss10,Biss09}. One might speculate that, rather than calculating spatially averaged quantities, a look at the aforementioned clusters in individual disorder realization itself would tell us more about the actual state the system is in. In this paper we propose a method to identify the different 
phases of the disordered BHM on the basis of geometric properties of what we call SF clusters, which are clusters of sites with non-integer boson occupation number. The MI phase then is the region in the phase diagram where no SF clusters exist, and the SF phase the region, where SF clusters percolate - the BG phase is in between: SF clusters exist, but do not percolate. We apply this criterion to results of LMF calculations and compare it with predictions of other methods: on one side SMF theory, where the individual phases are identified on the basis of spatially averaged LMF quantities like SF order parameter or compressibility, and on the other side QMC simulations, which are supposed to be exact up to statistical and extrapolation ($L\to\infty$, $T\to0$) errors (which can, of course, be quite large). Our aim is to demonstrate that the use of averaged quantities in LMF theory leads to incorrect predictions and that the cluster analysis predicts the phase diagram in $d\ge2$ in very good agreement with the 
exact QMC results, even when applied to LMF data.

The paper is organized as follows: In section \ref{section:model} we recapitulate the LMF and SMF theory for the disordered BHM. In section \ref{section:Criterion} we critically examine the use of the averaged SF order parameter $\overline{\psi}$ and the compressibility $\kappa$ as indicators of the different phases of the disordered BHM in LMF and SMF theory and then introduce our new method to construct the phase diagram on the basis of an analysis of SF clusters. We apply this cluster analysis in section \ref{section:Results} to the 2d disordered BHM with commensurate filling and with fixed disorder strength and compare it with prediction from QMC simulations and from SMF theory. Section \ref{section:Conclusion} concludes with a discussion of the equivalence of the predictions of the SF cluster analysis for the phase boundaries with the conventional definition of the MI-BG and BG-SF transition points.

\section{The model}\label{section:model}

Ultracold bosonic atoms in an two dimensional square optical lattice can be described by the BHM
\begin{equation}\label{BHM}
  \O{H}=\sum_i \left(\epsilon_i-\mu\right) \O[i]{n}+\frac{U}{2} \sum_i \O[i]{n} \left(\O[i]{n}-1\right)-J \sum_{\langle i,j\rangle} \Op[i]{a} \O[j]{a},
\end{equation}
where $i=1,\ldots,M$ is the site index, $M=L^2$ the number of sites
and $L$ the lateral size of the square lattice. The chemical potential
is described by $\mu$, the inter particle repulsion by $U$ and the
tunnelling rate by $J$. The last sum runs over all nearest neighbour
pairs $(ij)$ of the underlying lattice. The operator
$\O[i]{n}=\Op[i]{a}\O[i]{a}$ is the particle number operator of bosons
on site $i$, which are annihilated and created by the operators
$\O[i]{a}$ and $\Op[i]{a}$. Moreover, on-site disorder is introduced
by the parameter $\epsilon_i$, which is drawn randomly from a box distribution
$p(\epsilon)=\Theta\left(\Delta/2-|\epsilon|\right)/\Delta$, where
$\Delta$ is the strength of the disorder.

In the SF regime tunnelling dominates the system. Thus, the
ground state is a coherent state, which is an eigenstate of the
tunnelling part of the Hamiltonian. The SF parameter $\psi_i=\langle
\O{a_i} \rangle$ is the expectation value of $\O[i]{a}$ evaluated in
the ground state for $T=0$, which is non-zero for a coherent state. An
eigenstate of the diagonal part of the Hamiltonian on the other hand
is a Fock-state, which is the ground state of the system for small
tunnelling rates in the MI regime. The expectation value of $\O[i]{a}$
in a Fock-state is zero in any case. Hence, the mean value of the SF
parameters $\overline{\psi}=\sum_i \psi_i/M$ is an order parameter for
the SF phase.

The phase diagram of the pure system in dependence of $J Z / U$ and
$\mu / U$ consists of so called Mott lobes, in which the system is a
MI with a fixed integer number of atoms per site. While in the MI
regime the state is localized in real space, in the surrounding SF
regime it is localized in $k$-space. When disorder is introduced, the
Mott lobes shrink and a new phase, the BG phase, occurs. In order to
distinguish all three phases the
compressibility $\kappa=\langle \O{n}{}^2\rangle-\langle
\O{n}\rangle^2$ is necessary. Among the three phases only the MI is
non-compressible and only in the SF phase the SF order parameter is
non-zero. Consequently, the phase of the system can be identified
by the SF order parameter $\overline{\psi}=\sum_i \psi_i/M$ and the
compressibility $\kappa$. While the SF order parameter is a measure
for the coherence in the system, the compressibility describes the
variance of the particle number per site.

In order to determine the phase diagram of the BHM, the
ground state properties of the LMF Hamiltonian can be studied via SMF
theory \cite{Biss09,Biss10}, which computes the PD self-consistently,
or via LMF theory \cite{Shes93,Buon04b,Buon07}, which solves the
coupled set of equations for the local SF parameter directly on the
lattice. We will first describe the approximations made in LMF theory,
followed by a discussion of the additional assumptions made in the SMF
approach.

\subsection{Local mean field theory}\label{LMFtheory}

In LMF theory the tunnelling part of the Hamiltonian can be approximated via
\begin{equation}
  \O[i]{a} \Op[j]{a}\approx \O[i]{a} \langle \Op[j]{a} \rangle + \Op[j]{a} \langle \O[i]{a} \rangle-\langle \O[i]{a} \rangle \langle \Op[j]{a} \rangle,
\end{equation}
where terms of the form $(\O[i]{a}-\langle \O[i]{a} \rangle)
(\Op[j]{a}-\langle \Op[j]{a} \rangle)$ are neglected \cite{Shes93}. The central quantities are the local SF order parameters
\begin{equation}\label{SelfEqn}
 \psi_i=\langle gs | \O[i]{a} | gs\rangle,
\end{equation}
which are defined as the expectation values of the annihilation
operator at the individual site $i$ in the ground state of the
system. Because of the $U(1)$-symmetry they can be chosen to be
positive and real, which leads to
\begin{equation}\label{LMF}
  \O[i]{a} \Op[j]{a}\approx \psi_j \O[i]{a}+\psi_i \Op[j]{a}-\psi_i \psi_j.
\end{equation}
Thus, the Hamiltonian can be decomposed in a sum of diagonal operators,
\begin{eqnarray}\label{HLMF}
  \O{H}&=&\sum_i \O[i]{H}, \nonumber\\
  \O[i]{H}&=&\left(\epsilon_i-\mu\right) \O[i]{n}+\frac{U}{2} \O[i]{n} \left(\O[i]{n}-1\right)-J\eta_i \left(\O[i]{a}+\Op[i]{a}-\psi_i\right),
\end{eqnarray}
whose tunnelling rate is replaced by an effective local rate $J\eta_i$,
which depends on the local SF parameter of the neighbouring sites
$\eta_i:=\sum_j A_{ij} \psi_j$, with $A_{ij}=1$ for $i$ and $j$ nearest neighbours on the square lattice with periodic boundary conditions and zero otherwise. This approximation reduces the full
quantum problem to $M$ quantum sites, which are coupled in a mean
field way with a spatially varying coupling rate.

In order to compute the phase diagram in LMF theory the coupled set of the self-consistency equations 
\begin{equation}\label{self-consistent}
  \psi_i=\langle \O[i]{a} \rangle, \quad i=1,\ldots,M, \quad M=L^2
\end{equation}
is solved on a $L \times L$ lattice, where the expectation value is
evaluated in the ground state of $\O[i]{H}$ that itself depends an the
local SF parameter $\psi_i$. As a result of the \mbox{decomposition
\eqref{HLMF}} all states considered (in particular the ground state) are a direct product of individual single-site states. This means in particular that they are Gutzwiller states of the form 
\begin{equation}\label{Gutzwillerstate}
 |\Psi \rangle = \prod_{i=1}^M\left( \sum_{n=0}^\infty c_n^i |n\rangle_i\right).
\end{equation}
with single-site states $|\phi_i\rangle= \sum_{n=0}^\infty c_n^i |n\rangle_i$ given in particle number basis and $|c_n^i|^2$ describes the probability to find $n$ bosons at site $i$ and fulfils $\sum_{i=1}^M |c_n^i|^2=1$. In particular, the local SF order parameter is then given by $\psi_i=\sum_{n=0}^\infty {c_{n-1}^i}^* c_n^i \sqrt{n_i}$ and the local boson number is represented by $\langle \O[i]{n} \rangle=\sum_{n=0}^\infty |c_n^i|^2 n_i$.

For the numerical implementation, starting from a random initial configuration for $\psi_i$ on the 2d lattice, the set
\eqref{self-consistent} of equations is solved recursively. This involves solving the eigenvalue problem on each site and computing the expectation value of the annihilator in the numerically determined ground state. This is repeated until the
averaged SF order parameter
\begin{equation}
\overline{\psi}=\left[\frac1M \sum_{i=1}^M{\psi_i}\right]_{\rm av}
\end{equation}
is determined with an accuracy of $10^{-4}$. In the disordered case
the results are averaged over $200$ different realizations of disorder,
indicated by the brackets $[\ldots]_{\rm av}$. Since we
are working in a regime in which the maximum average particle number per site is three, it was numerically checked that it is sufficient to truncate the
basis of the Hilbert space for each site at $N=10$. With the solution
found for the local SF parameters $\psi_i$ on the lattice, the
ground state of Hamiltonian \eqref{HLMF} is calculated numerically. Afterwards all desired expectation values and finally the
compressibility
\begin{equation}
\kappa=\left[\langle
\O[]{N^2}\rangle-\langle\O[]{N}\rangle^2\right]_{\rm av},
\end{equation}
with $\O[]{N}=\sum_i\O[i]{n}$ can directly be computed. The probability distribution (PD)
\begin{equation}
P(\psi)=\left[\frac1M \sum_{i=1}^M \delta(\psi-\psi_i)\right]_{\rm av}
\end{equation}
is determined on the basis of the complete set
of values of $\psi_i$, additionally averaged over disorder
realizations.

\subsection{Stochastic mean field theory}\label{subsection:SMF}

The central idea of SMF theory is to circumvent the computation of all
local order parameters $\psi_i$ by deriving a self-consistency
equation for the probability distribution $P(\psi)$
directly \cite{Biss09, Biss10}. Additional approximations are of course necessary. The SF
order parameter $\psi=\langle gs | \O[]{a}| gs\rangle$, which is
derived from the ground state of the full quantum Hamiltonian, can be
determined by the ground state of the single-site Hamiltonian
\eqref{HLMF} in dependence of the stochastic variables $\epsilon$ and
$\eta$. The parameter $\epsilon$ is then a stochastic variable drawn
from the disorder distribution $p\left(\epsilon \right)$ and as a
result $\psi=\langle gs | \O[]{a} | gs\rangle$ is a stochastic
variable, drawn from the PD $P \left(\psi\right)$, which must be
determined self-consistently. Since $\eta$ is the sum of the SF
parameters of the neighbouring sites it also is a stochastic variable
drawn from $Q\left( \eta \right)$. The problem of computing the
ground state of the full quantum system for all lattice sites
simultaneously is thereby replaced by analysing the ground state $| gs
\left(\epsilon, \eta \right)\rangle$ of the site independent
Hamiltonian
\begin{equation}\label{HSMF}
  \O[]{H}=\left(\epsilon-\mu\right) \O[]{n}+\frac{U}{2} \O[]{n} \left(\O[]{n}-1\right)-J\eta \left(\O[]{a}+\Op[]{a}-\psi \right)
\end{equation}
as a function of $\epsilon$ and $\eta$. Thus, the probability distribution (PD)
\begin{equation}\label{SelfEqnP}
P \left(\psi \right)= \int \d \eta Q\left( \eta \right) \tilde P_\eta \left(\psi \right)
\end{equation}
depends on the distribution $Q\left( \eta \right)$ of the occurring
values of $\eta$ and the distribution $\tilde P_\eta
\left(\psi\right)$ of the local SF parameters for given $\eta$. A
direct analysis of $\langle gs\left(\epsilon, \eta
\right)|\O[]{a}|gs\left(\epsilon, \eta \right)\rangle$ as a function
of $\epsilon$ and $\eta$ yields
\begin{equation}
\tilde P_\eta \left(\psi \right)= \frac{\d}{\d \psi}\int \d\epsilon p\left( \epsilon \right) \Theta \left( \psi-\langle gs(\epsilon,\eta)| \hat a | gs(\epsilon,\eta)\rangle\right).
\end{equation}
Since $\eta$ is the sum of the local SF parameters $\psi$ of the
neighbouring sites its distribution is given by
\begin{equation}\label{SMFQ00}
Q\left( \eta \right)= \int_0^\infty \prod_{i=1}^Z \d\psi_i \mathcal P_Z \left( \psi_1,\ldots,\psi_Z \right) \delta
\left(\eta-\sum_{i=1}^Z \psi_i \right),
\end{equation}
where $P_Z \left( \psi_1,\ldots,\psi_Z \right)$ is the connected
probability distribution function of the local order parameters
$\psi_1,\ldots,\psi_Z$ of the $Z$ neighbour sites of a
single-site. Assuming that these $Z$ local SF parameters are
statistically independent
\begin{equation}\label{SMFRestriction}
 \mathcal P_Z\left(\psi_1,\ldots,\psi_Z\right)= 
\prod_{i=1}^Z P\left(\psi_i\right),
\end{equation}
equation (\ref{SMFQ00}) transforms into a convolution
\begin{equation}\label{SMFQ}
Q\left( \eta \right)= \int_0^\infty \left( \prod_{i=1}^Z \d\psi_i P \left( \psi_i\right) \right) \delta \left(\eta-\sum_{i=1}^Z \psi_i \right),
\end{equation}
Since the assumption (\ref{SMFRestriction}) implies the absence of
correlations of the local SF parameters, one expects that it is not
justified close to the phase boundaries, where the correlation length
even diverges, when the transition is 2nd order. We examine the validity of this approximation in
dependence of the system parameter $J Z/U$ and $\mu/U$ in \ref{section:assumptionSMF}.

After determining the PD $P\left(\psi \right)$ the SF order parameter
$\overline{\psi}=\int\d\psi P\left( \psi \right) \psi$ is given by the
mean value of the distribution. The compressibility
\mbox{$\kappa=[\langle \O[]{N^2}\rangle-\langle\O[]{N}\rangle^2]_{\rm av}$} 
with $\O[]{N}=\sum_i\O[i]{n}$ is computed. With these quantities at
hand one can, on the basis of the underlying approximations, estimate
the phase boundaries of the transitions between MI, BG and SF. They are shown in Figure \ref{PhaseClu} and \ref{PhaseSMFClu} and will be discussed in the following.

\section{Criterion for the phase transition}\label{section:Criterion}

In this section we will first discuss the well known SF order parameter $\overline{\psi}$ and the compressibility $\kappa$, which are expected to indicate the phase transition: The ground state in the MI regime is a Fock-state, which is incompressible $\left(\kappa=0\right)$ and non-coherent $\overline{\psi}=0$, while conversely in the SF regime it is described by a coherent state $\overline{\psi}=0$, which is compressible $\left(\kappa>0\right)$. If disorder is introduced, those phases are separated by the BG phase, which is compressible $\left(\kappa>0\right)$, but  not coherent $\overline{\psi}=0$. We will see that a precise prediction of the transition point on the basis of $\overline{\psi}$ or $\kappa$ is not possible in LMF theory. Instead we will introduce an identification criterion of the different phases on the basis of the complete set of local occupation numbers.

\subsection{SF order parameter and compressibility}\label{section:LSPC}
\begin{figure}[!htb]
\begin{center}
\includegraphics[width=15cm]{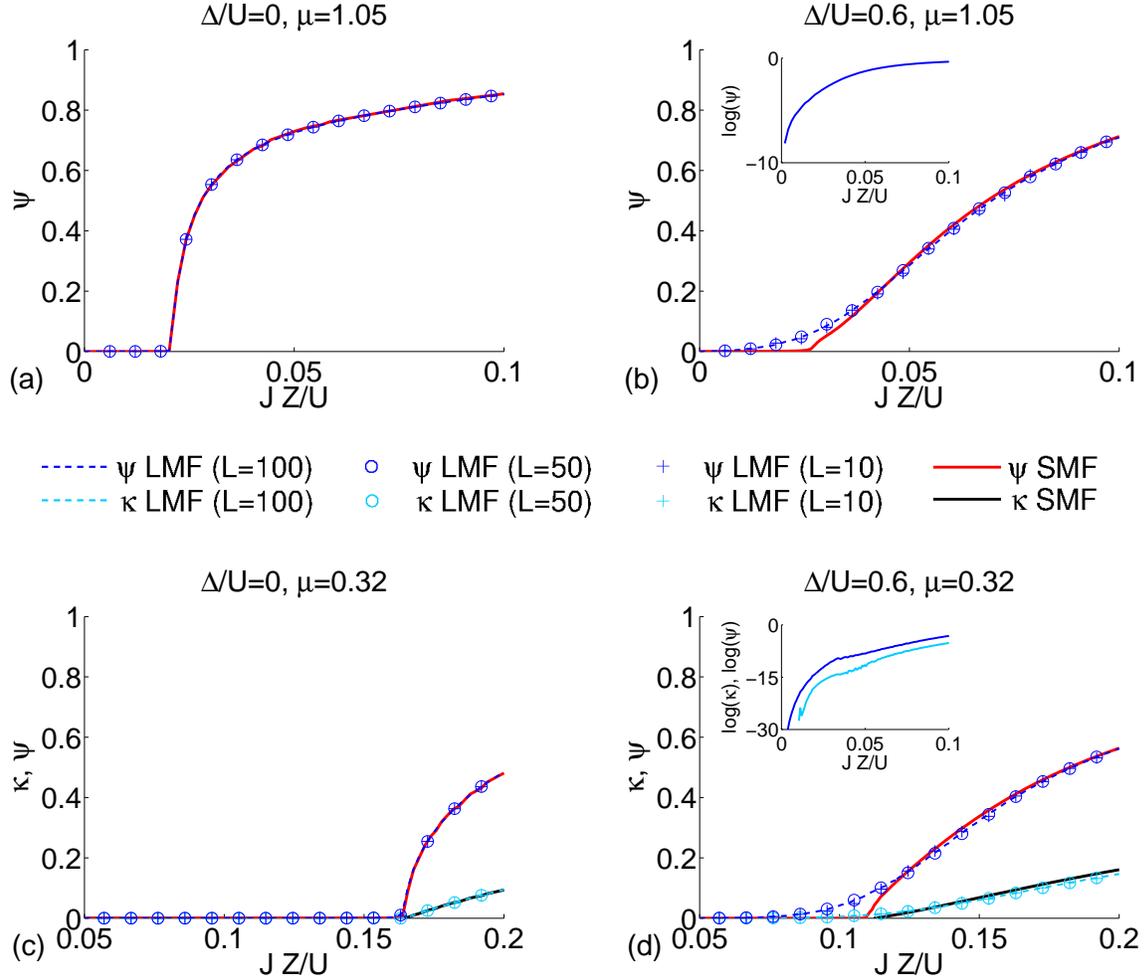} \caption{\label{SchnittM2} 
Comparison of the LMF and SMF predictions for the average SF order
parameter $\overline{\psi}$ and the compressibility $\kappa$ for fixed
chemical potential $\mu$ and varying tunnelling rate $J$.  {\bf Left:}
Homogeneous case ($\Delta=0$), {\bf Right:} disordered case with
$\Delta/U=0.6$. Top row is for $\mu=1.05$, where the ordered system
displays a MI SF transition and the disordered system a BG SF
transition ($\kappa>0$ for all values of $J$).  Bottom row is for
$\mu=0.32$, where the ordered system again displays a MI SF transition
and the disordered system is expected to display MI, BG and SF phases
(see section 3.2). For LMF theory the results for a 2d lattice with
$L=100$ (line), $L=50$ ($\circ$), $L=10$ ($+$) are depicted, which
shows that finite-size effects can be neglected.}
\end{center}
\end{figure}

In the ordered case the on-site energies $\epsilon_i$ are zero and the
lattice is homogeneous. The SF order parameter $\overline{\psi}$
clearly marks the location of the MI to SF phase transition as shown
in Figure \ref{SchnittM2} on the left hand side for $\mu/U= 1.05$ and $0.32$. While the SF
order parameter is zero for small tunnelling rates in the MI
phase, it becomes non-zero and positive above a critical value of the
tunnelling rate in the SF phase. The compressibility $\kappa$ shows the same
behaviour at the phase transition as the SF order parameter for both
methods in the ordered case. Moreover, we analysed different
lattice sizes $L$ and the LMF results show no visible finite-size
effects. In this way the phase transition in the ordered case can be
determined very precisely, both within SMF and LMF theory. The
resulting phase transitions agree perfectly with the perturbation predictions
\cite{Oost01}
\begin{equation}\label{perturbative}
 \frac{\mu}{U}=-\frac{1}{2}\left(\frac{J Z}{U}-2 n+1\right)\pm \sqrt{\frac{1}{4}\left(\frac{J Z}{U}-1\right)^2-\frac{J Z}{U} n}\,,
\end{equation}
where $n$ denotes the mean number of particles per site and
simultaneously counts the number of lobes. The calculation in
\cite{Free96} predicts that in the disordered case the upper (lower)
part of the Mott lobes are shifted downwards (upwards) by $\Delta/2$
but the shape remains unchanged. 

The situation for the disordered case is shown in Figure
\ref{SchnittM2} on the right hand side. The SF order parameter is 
shown for $\mu/U= 1.05$ and $0.32$ as a function of the tunnelling rate
for a disorder of $\Delta/U=0.6$. Whereas the SMF theory predicts a
direct BG SF transition, at a critical value $JZ/U\approx0.0241$ and
$1.0455$, see \ref{SchnittM2} (b, d), above which $\overline{\psi}$
become non-zero the behaviour of $\overline{\psi}$ as predicted by LMF
theory does not indicate a transition at all; it varies smoothly with
the tunnelling rate $J$.  This is not a finite-size effect as we have
checked by examining different lattice sites, as shown in
\ref{SchnittM2}. The compressibility, which indicates the MI BG
transition, displays the same behaviour.

It turns out that the reason for the failure of the average SF order
parameter to predict the location of the BG SF boundary is
the following: In the disordered case the value of the local SF
parameter varies substantially from site to site due to the
variation of the local potential of $\epsilon_i$. Close to the phase
transition there are sites with zero local SF parameter and others,
where the local SF parameter is still positive. This has been
interpreted as an overestimation of the phase coherence in LMF
description \cite{Biss09}. Our interpretation, however, is different:
It is only the average SF order parameter $\overline{\psi}$ that
overestimates the phase coherence. A closer look at the complete
probability distribution $P(\psi)$ of the local SF order parameters and
their geometrical features provides an estimate of the SF
regions in the phase diagram. Its prospects are discussed in the \ref{section:PD}. In the next section we discuss, how a deeper understanding of the mechanisms driving the phase transitions and their location in the phase diagram can be obtained by studying the geometric characteristics of the spatial inhomogeneities of the local SF parameters $\psi_i$ and particle number per site $\langle\O[i]{n}\rangle$.

\subsection{Identification of phases via local boson occupation number}\label{section:PhasesLMF}

The MI and SF phases can be discriminated by the boson number
statistics at individual sites, as has also been demonstrated
experimentally in \cite{Grei02}. The ground state in
the extreme MI limit $(J\to0)$ is a Fock states with a
definite number of particles $n$ at each site. In the extreme SF limit
$(U\to0)$, the ground state is a coherent state , in which the local
boson number distribution is close to a Poissonian. Although, in the
regime between these two extreme limits the ground state wave function
can no longer be written as simple product states still the MI phase
is characterized by a sharp, integer boson number per site and the SF
phase by a fluctuating boson number per site, i.e. a non-vanishing
variance of the boson number distribution $p_n^i=|c_n^i|^2$ (c.f. the
expansion coefficient in the Gutzwiller wave function \eqref{Gutzwillerstate}). In other
words in the MI regime the expectation value of the number of bosons
per site $\langle\O[i]{n}\rangle$ is an integer, whereas in the SF regime it is
non-integer.

Whereas in the ground state of the homogeneous BHM
either all sites have an integer boson number (MI regime) or all sites
have a non-integer boson number (SF regime) this is different in the
disordered BHM. In particular, outside of the MI regime one
expects to encounter spatially inhomogeneous situations, in which some
sites have a sharp (integer) boson occupation numbers and others have
fluctuating (non-integer) boson numbers. Introducing phase operators
$\Phi$ that is canonically conjugate to the boson number operators
$\hat{n}_i$ the BHM can be mapped upon a Josephson
junction array or more general to a quantum rotor model \cite{Fish89}, in which
superfluidity is indicated by long range order in these phases
($d\ge2$). Because of the Heisenberg uncertainty relation sites with
sharp phases correspond to sites with fluctuating boson numbers, and
connected clusters of sites with fluctuating boson numbers tend to
have, roughly speaking, all the same phase. These clusters can
therefore be identified with SF regions, although, true
superfluidity only exists in the infinite system. Indeed, once these
phase ordered clusters percolate, true superfluidity emerges,
signified by a non-vanishing SF stiffness, which is the extra
free energy cost to impose a uniform twist on the phases. Since such a
uniform twist can be introduced by applying a certain twist at the
boundary phases in one space direction, it is clear that the SF
stiffness is zero as long as the clusters do {\it not} percolate: In
the absence of long range order in the phases such a twist at the
boundary does not cost energy.

On the basis of this qualitative picture we hypothesize that the BG to
SF transition in the $d$-dimensional BHM ($d\ge2$)
coincides with the percolation transition of connected clusters of
lattice sites with a non-integer boson number expectation value
$\langle\O[i]{n}\rangle$. We expect this coincidence to hold as long as the
SF phase displays true long range order, characterized within the
phase description by a non-vanishing long distance limit of phase
correlations - which means it should hold for $d\ge2$. The BG SF
transition in one dimensional BHM might not be related
to a percolation transition, since in $d=1$ the SF phase has only
quasi long range order (algebraically decaying correlations).
We should note that the relation between the BG SF transition and
percolation has already been pointed in \cite{Shes95,Buon09,Dell11}, but has neither
been used in a quantitative manner to determine phase boundaries
nor checked against, for instance, Monte Carlo results.
\begin{figure}[!htb]
\begin{center}
\includegraphics[width=15cm]{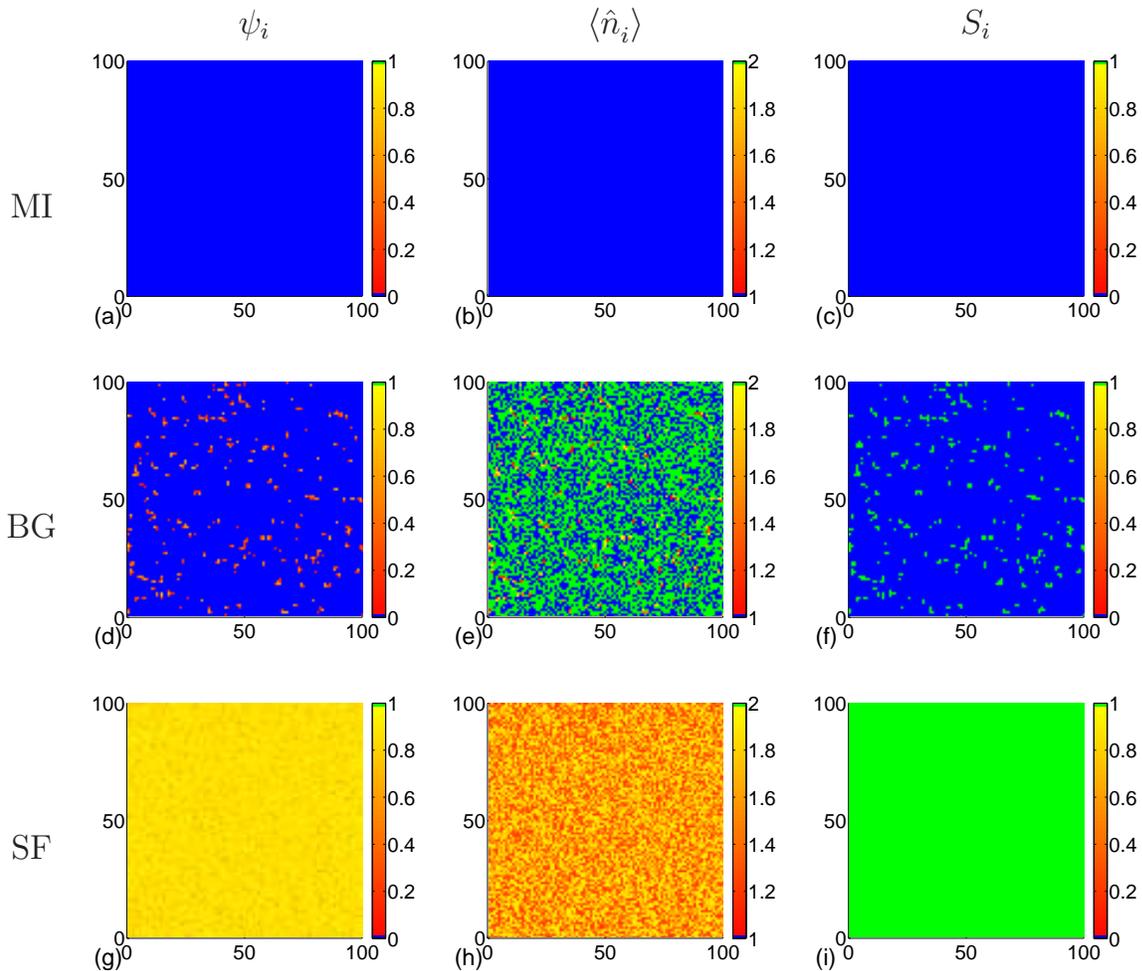}
\caption{\label{Lattice} Configurations of the local SF parameter $\psi_i$, the occupation number $\langle\O[i]{n}\rangle$ and the discrete variable $S_i$ \eqref{discreteMap} for a single realization of disorder for $\Delta/U=0.6$. The first row shows an example for MI ($JZ/U=0.0242,\;\mu/U=0.4394$) followed by one for BG ($JZ/U=0.0182,\; \mu/U=1.0455$) and SF ($JZ/U=0.141,\; \mu/U=1.0455$). Note that blue marks the minimal value (zero in the left and right, one in the middle column) and green the maximal value (one in the left and right, two in the middle column).}
\end{center}
\end{figure}

In the following we denote the sites with non-integer boson occupation
number $\langle\O[i]{n}\rangle$ as SF sites, and sites with integer $\langle\O[i]{n}\rangle$ as MI sites. Analogously, we discriminate SF clusters and
MI clusters. Formally we map the boson occupation numbers to a
discrete field $S_i$ that is set to $S_i=1$ for SF sites and $S_i=0$
for MI sites. Then we identify the different phases of the disordered
BHM as follows:

MI phase: $S_i=0$ for all sites $i$. All boson occupation numbers are
integer (and identical), consequently the compressibility $\kappa$ is
zero.

SF phase: $S_i=1$ for a macroscopic fraction of sites, which form a
percolating connected cluster. According to what we discussed above the
percolating cluster has phase long range order and thus yields a
non-vanishing SF stiffness (which is proportional to the
SF density).

BG phase: Characterized by a non-vanishing density of sites with
$S_i=1$, none of the connected clusters formed by the SF sites
percolates. The BG phase is thus characterized by isolated SF clusters within a MI sea. The phases of the isolated clusters are
uncorrelated, hence phase long range order is lacking and the
SF density vanishes (no SF order). Moreover, due to the
number fluctuations on the SF sites the BG phase is compressible
($\kappa>0$).

Within LMF theory the expectation values of the local boson
occupation numbers are straightforward to calculate via
$n_i = \langle gs|\O[i]{n}|gs \rangle$, where$|gs\rangle$ is the ground state of the
LMF Hamiltonian \eqref{HLMF}. For numerical reasons we introduce a
threshold $\gamma_n$ into the definition of the discrete field
\begin{equation}\label{discreteMap}
S_i= 
\begin{cases}
0 &\text{if}\quad I-\gamma_{n} \leq \langle\O[i]{n}\rangle \leq I+\gamma_{n}\quad I=0,1,2,\ldots\text{ ,}\\
1 & \text{else,}
\end{cases}
\end{equation}
where $\gamma_n=5\cdot10^{-3}$ is chosen to serve as the cut-off in this algorithm. In the whole parameter range, where sites with integer particle number occur, the histogram of the mean particle number $\langle \O[i]{n}\rangle$ has narrow peaks of width $\gamma_n$ at integer values. The width decreases when we increase the number of iteration steps to solve the self-consistency equations \ref{self-consistent}. The threshold parameter $\gamma_n$ introduced to identify MI sites (and complementarity SF sites) can be reduced by increasing the numerical effort without changing the final results. 

In Figure \ref{Lattice} typical results for one realization of disorder for $\Delta/U=0.6$ and $\mu/U=1.0455$ are shown for three different values of the tunnelling rate $J Z/U$ for the three phases . In the first row the local SF
parameter $\psi_i$, in the second the particle number per site
$\langle \O[i]{n} \rangle$ and in the third the resulting discrete map
$S_i$ is shown. In the MI regime all sites are occupied by the same integer number of particles (in this case one, since we are in
the first Mott lobe). At the transition from the MI to the BG regime SF sites $\left(S_i=1\right)$ with non-integer particle number occur. Because of these locally occurring SF sites $\psi_i>0$ the SF order parameter $\overline{\psi}$ is small but not zero in this regime. Since the SF islands are compressible, this phase has positive compressibility. In the BG phase the SF islands does not percolate, yet. They grow in number and size, until one of them finally percolates. The percolation represents the actual transition to the SF regime in parameter space. Just after the percolation the phase 
in the system is coherence macroscopically, which means that all local SF parameter $\psi_i$ are positive and compressible, as described above.

\subsection{Percolation analysis}\label{section:Percolation}
\begin{figure}[!htb]
\begin{center}
\includegraphics[width=15cm]{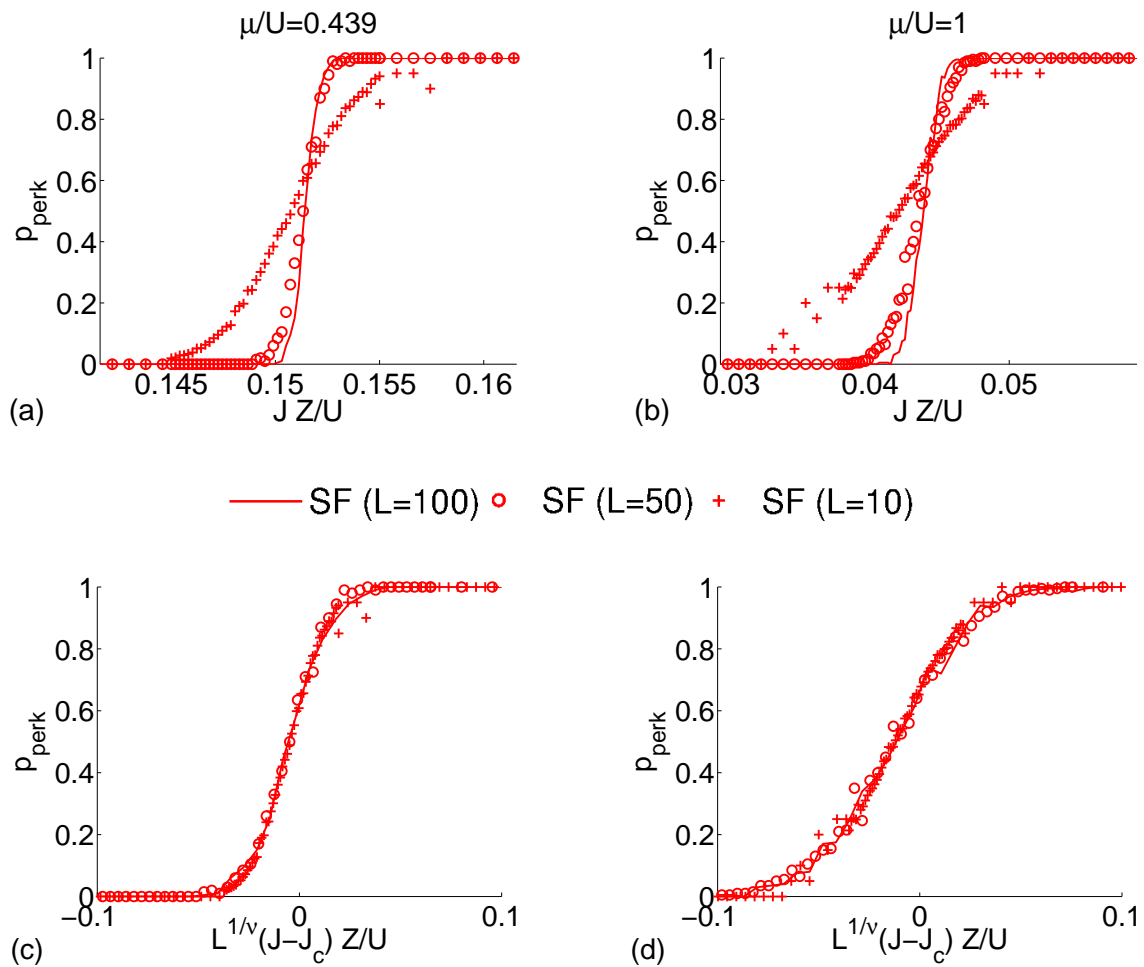}
\caption{\label{clusterPerk}The percolation probability $p_{\rm{Perc}}$ of the SF cluster (\textbf{top}) and finite size scaling plot (\textbf{bottom}). The critical tunnelling rate according to the finite-size scaling \eqref{FiniteSize} are given by $J_c Z/U=0.15$ for $\mu/U=0.439$ (\textbf{left}) and $J_c Z/U=0.04$ for $\mu/U=1$ (\textbf{right}); the critical exponent is $\nu=1.33$.}
\end{center}
\end{figure}
In this section we demonstrate how we determine numerically the percolation transition via a cluster analysis of the discrete map $S_i$ and finite size scaling \cite{Stau94}. Assume we study the phase diagram in dependence of the system parameter called $x$ and $y$. Than, the percolation probability $p_{\rm{Perc}}$, i. e. the probability of having a percolating cluster, is given for fixed $y$ as a function of $x$ and will be determined for different system sizes $L$. The percolation probability is expected to obey the finite-size scaling form
\begin{equation}\label{FiniteSize}
p_{\rm{Perc}}\left(L,x\right) = \tilde p\left(L^{1/\nu}\left(x-x_c\right)\right),
\end{equation}
where $x_c$ is the percolation threshold, i.e.\ the value above which a percolating cluster exists with probability one, and
$\nu$ the critical exponent determining the divergence of the mean
lateral cluster size at the transition. The scaling function $\tilde p(X)$
approaches zero for $X\ll1$ and one for $X\gg1$, which means that
exactly at the transition $x_c$ the curves for different system size
should intersect (in the scaling limit). This intersection point,
which we can easily be identified with the system sizes behaviour at hand, is
thus a reliable indicator for the percolation transition.

The cluster analysis of the discrete map $S_i$ is done for every disorder realization. Afterwards the results are averaged over $200$ ($L=50,100$) and $2500$ ($L=10$) realizations of disorder. The percolation probability $p_{\rm{Perc}}$ for this case is shown in Figure \ref{clusterPerk} (blue for MI and red for SF sites) for different system sizes as a function of the tunnelling rate $J Z/U$. Moreover, the finite size scaling analysis for the percolation transition at $\mu/U=0.439$ and $\mu/U=1$ for $\Delta/U=0.6$, yielding $J_c Z/U=0.15$, $J_c Z/U=0.04$ respectively and the critical exponent $\nu=1.33$ in both cases, is depicted. Thus, this transition is in the universality class of conventional 2d percolation \cite{Stau94}. We find the same universality class of the percolation transition for all parameter values that we studied.

\section{Results}\label{section:Results}

\subsection{Commensurate filling - Comparison with QMC results}\label{section:Commensurate filling}
In this section we determine the complete phase diagram for commensurate density in dependence of $\Delta/2 J$ and $U/J$, for which a prediction on the basis of QMC simulations is available \cite{Soey11}. We fix the particle density to $\overline{n}=\langle \sum_{i=1}^M \O[i]{n} \rangle/M=1$ with an accuracy of $10^{-4}$ by adjusting the chemical potential for each point ($\Delta/2 J$, $U/J$) in the phase diagram that we study. Outside of the Mott lobes this result is unique, whereas in the MI regime the chemical potential is fixed to the middle of the MI gap. In the $\mu/\Delta$ versus $JZ/U$ representation, where the Mott lobes are visible and which we will discuss in section \ref{section:fixed Disorder}, this $\overline{n}=1$ line always passes the tip of the first Mott lobe. In the $\Delta/2J$ versus $U/J$ parameter space the corresponding line for fixed $\Delta$ is a straight line through the origin with slope $\Delta/2 U$.

\begin{figure}[!htb]
\begin{center}
 \includegraphics[width=15cm]{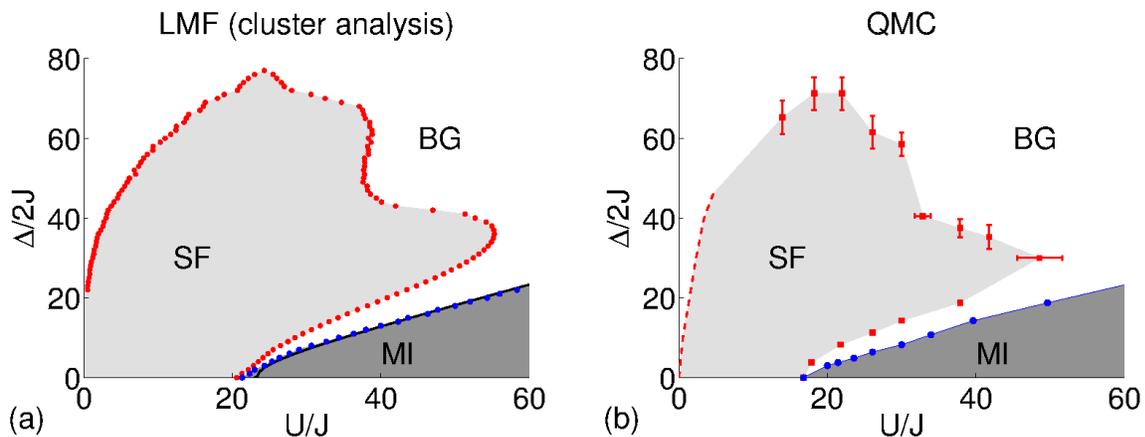}
 \caption{\label{PhaseClu} {\bf Left:} LMF cluster analysis phase diagram for commensurate boson density  $\overline{n}=\langle \sum_{i=1}^M \O[i]{n} \rangle/M=1$ determined with the discretized boson occupation number field $S_i$ \eqref{discreteMap}. The percolation transition of the SF sites $(S_i=1)$ occurs when crossing the red line, which indicates the BF-SF phase boundary. The blue line marks the boundary of the MI region, in which all sites are MI sites $(S_i=0)$. The black line indicates the MI-BG transition according to the perturbative result \eqref{perturbative}. {\bf Right:} Prediction for the phase diagram for commensurate boson density $\overline{n}=1$ based on the results of QMC simulations (data taken from \cite{Soey11}).}
\end{center}
\end{figure}

With the chemical potential that fixes the density $\overline{n}$ to one
we compute the ground state of the LMF Hamiltonian \eqref{HLMF} and determine
the discretized boson number field $S_i$ \eqref{discreteMap}, which we use to identify
MI, BG and SF phase. The resulting phase diagram is shown in Figure \ref{PhaseClu} on the left. As expected \cite{Poll09} the SF region is
completely surrounded by the BG phase (except at $\Delta=0$). Its boundary
has some characteristic features: It extends in a slight bump up to
quite large disorder strength (up to $\Delta/2J\sim75$) and in a pronounced
nose up to the interaction strength $U/J\sim52$. This nose gives rise to a
re-entrant behaviour: Moving vertically from a point within the MI
phase, which has long range positional order, one enters first the BG
phase, which is disordered and then, upon further increasing the
disorder strength, enters the SF phase, which has off-diagonal long
range order. Weak disorder thus supports superfluidity in the
BHM, as has been observed before \cite{Krau91,Soey11,Guar09,Lin11}.

Remarkably, our prediction on the basis of a cluster analysis of LMF ground states agrees very well with the results of
QMC simulations \cite{Soey11} shown for comparison in
Figure \ref{PhaseClu} on the right. The shape of the SF-BG phase boundary with its characteristic
nose and bumps clearly coincide. The
quantitative agreement is very good, too, regarding the substantial
error bars of the QMC data in the large disorder and large interaction
regime (the QMC estimate for the extreme value of $\Delta$ in the bump
is $\Delta/2J\sim72\pm4$ and of inter particle interaction in the nose $U/J=49\pm3$, c.f Fig. 2 in
\cite{Soey11}). Moreover, with our method we could also explore the weak
interaction region, which is hardly accessible by QMC methods, and found a
singular behaviour of $\Delta$ with $U\to0$, which is compatible with the
analytically predicted behaviour $\Delta\propto\sqrt{U}$ \cite{Falc09}.  We
conclude that the percolation criterion that we introduced in section \ref{section:PhasesLMF} to locate the SF-BG transition produces remarkably accurate
predictions even in LMF theory.

Our result for the MI-BG transition line, which denotes the
appearance of non-integer boson occupation numbers and thus SF sites,
agrees well with the perturbative result \ref{perturbative}, shown in Figure \ref{PhaseClu} on the left. Moreover, they agree with the line $\Delta=E_{g/2}$ obtained using the gap data from \cite{Capo08}, shown in Figure \ref{PhaseClu} on the right.

In passing, we note that for weak disorder the MI clusters percolate close to the BG-SF transition line, whereas for
stronger disorder they percolate deeper inside the BG phase. Whereas for weak disorder the individual sites of a MI cluster in the BG phase all have the same integer occupation number, this is in general not the case any more for strong disorder: the integer occupation number of MI clusters can vary from site to site.

Finally, we note that SMF theory as described in section \ref{section:LSPC} predicts a
direct MI-SF transition along the lower border of the SF region in the
parameter range shown in Figure 4. The characteristic BG region for
small disorder strength is absent in this parameter range, which is in
contradiction to the theorems proven in \cite{Poll09}, which exclude a direct
MI-SF transition in any disordered systems. Besides, we checked that
the BG phase occurs for even higher values of $U/J$, which is in
agreement with results to be presented in the next section. 

\subsection{Fixed Disorder - Comparison with SMF theory}\label{section:fixed Disorder}

After we have seen in the last section that our method to determine the phase diagram of the 2d disordered BHM leads to results that agree very well with QMC predictions, we determine in this section the ($\mu/U$-$J Z/U$) phase diagram for a fixed disorder strength $\Delta/J=0.6$ and compare it with predictions of SMF theory. In this phase diagram the Mott lobes occur and the line given by $\langle n \rangle=1$ always passes the tip of the first one.
\begin{figure}[!htb]
\begin{center}
 \includegraphics[width=15cm]{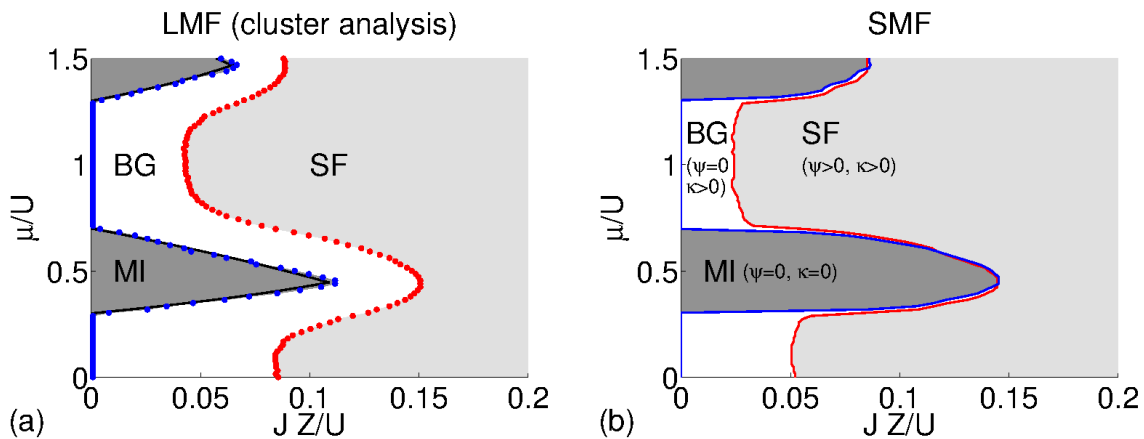}
 \caption{\label{PhaseSMFClu} Comparison of LMF cluster analysis and SMF phase diagram for fixed disorder strength $\Delta/U=0.6$. {\bf Left:} LMF cluster analysis phase diagram determined with the discretized boson occupation number field $S_i$ \eqref{discreteMap}. The percolation transition of the SF sites $(S_i=1)$ occurs when crossing the red line, which indicates the BG-SF phase boundary. The blue line marks the boundary of the MI region, in which all sites are MI sites $(S_i=0)$. The black line is the MI-BG transition according to the perturbative result given by \eqref{perturbative}. {\bf Right:} SMF phase diagram determined by using the SF order parameter $\overline{\psi}$ and the compressibility $\kappa$ \cite{Biss09, Biss10}. The red line indicates the critical tunnelling rate $J$ where the SF order parameter $\overline{\psi}$ becomes non-zero, the blue line the critical tunnelling rate, where the compressibility $\kappa$ becomes non-zero.}
\end{center}
\end{figure}
\begin{table}[!htb]
 \begin{tabular}{l|ll|ll|l}
 & \multicolumn{2}{c}{MI-BG} & \multicolumn{2}{|c|}{BG-SF}&\\
 & $J Z/U$& $\mu/U$& $J Z/U$& $\mu/U$&$\Delta/U$\\ \hline
LMF (cluster analysis) & $0.1115$& $0.4644$ & $0.1509$& $0.4434$&$0.6$\\
QMC results \cite{Soey11} & $0.124$& ($0.4561$) & $0.2012$& ($0.4082$)&$0.6$\\
strong-coupling expansion \cite{Free96}&$0.1253$&$0.4345$&&&$0.6$\\ \hline
LMF (cluster analysis) & & & $0.0942$& $0.4868$&$1$\\
QMC results \cite{Soey11} & & & $0.1047$& ($0.4846$)&$1$\\ \hline
LMF (cluster analysis) & & & $0.0934$& $0.5043$&$2$\\
QMC results \cite{Soey11} & & & $0.1062$& ($0.4950$)&$2$\\
quantum rotors model \cite{Lin11}&&&$0.112$&$0.375$&$2$
\end{tabular}
\caption{\label{quantumComparison} Comparison of the parameters at the tip of the first Mott lobe of quantum models \eqref{BHM} with LMF cluster analysis results. The chemical potential $\mu/U$ for the QMC results of \cite{Soey11} is our LMF estimate for a density $\overline{n}=1$ and fixed values of $JZ/U$ and $\Delta/2J$. The BG-SF predictions of \cite{Soey11} were not obtained by QMC of the disordered BHM, but are based on gap data for the ordered BHM \cite{Capo08}.}
\end{table}

In section \ref{subsection:SMF} we introduced SMF theory and already emphasized that SMF theory bases on the same approximation to the Hamiltonian as LMF theory, but, it involves the additional approximation \eqref{SMFRestriction} on the distribution $P_Z \left( \psi_1,\ldots,\psi_Z \right)$. The validity of this restriction fails close to the phase transitions as we show in \ref{section:assumptionSMF}. Despite or perhaps because of this approximation the SF order parameter $\overline{\psi}$ as well as the compressibility $\kappa$ computed within SMF theory are exactly zero in specific regions of the parameter space (c.f. Figure \ref{SchnittM2}), which one might want to identify with MI and BG phase, as was done in \cite{Biss09, Biss10}, c.f. section \ref{section:LSPC}. The LMF cluster analysis and SMF \cite{Biss09, Biss10} phase diagrams for fixed disorder strength $\Delta/U=0.6$ are shown in Figure \ref{PhaseSMFClu}. One immediately observes substantial differences: Firstly, in LMF theory the BG phase 
always 
separates the MI from the SF phase. The intervening BG phase is actually predicted to be quite large even at the tip of the Mott lobes, not just a "thin sliver" \cite{Fish89}. SMF theory, however, predicts a direct MI-SF transition, in contradiction to \cite{Poll09}. Secondly, large differences in the critical tunnelling rate for the BG-SF transition occur especially in the region around $\mu=1$. Assume we fix the chemical potential there. In this case the SMF theory predicts the phase transition at $J Z/U=0.0241$. The percolation of the SF cluster, however, takes place at $J Z/U=0.0430$. Thus, significant changes of the system in this case occur for values of the tunnelling rate twice as large as predicted by $\overline{\psi}$ in SMF theory.

A direct comparison of our results with the QMC data of \cite{Soey11} is not possible here, since the latter are obtained for the canonical ensemble, where the chemical potential is absent. However, we can take our LMF estimate for the value of $\mu$ that fixes the particle density at $\overline n=1$ for fixed $U/J$ and $\Delta/2J$ to obtain an approximate comparison - see Table \ref{quantumComparison}, where we also show the prediction of the strong coupling expansion \cite{Free96} for the MI-BG transition for $\Delta=0.6$. One observes deviations of the QMC and strong coupling predictions from our LMF cluster analysis results at the tip of the Mott lobe, but a good agreement for stronger disorder, $\Delta/U=2$. At this disorder strength also a QMC prediction for the quantum rotor model exists \cite{Lin11}, which differs by $25\%$ from the predictions for the BHM, ours and the one from QMC. We also note that the tapered shape of the Mott lobe predicted by the strong coupling expansion \cite{
Free96} agrees well with our result of the LMF cluster analysis shown in Figure \ref{PhaseSMFClu}.

In addition to our LMF cluster analysis and the SMF theory discussed above a number of other approximative methods have been applied to calculate the phase diagram of the BHM at fixed disorder. In the zero temperature mean field phase diagram of \cite{Krut06} the BG phase is completely absent, which might be true in infinite dimensions, but certainly not in $d=2$ or $3$.

A LMF theory has been used in \cite{Buon07,Buon09} to solve the self consistency equations \eqref{SelfEqn} and to calculate a LMF expression for the stiffness or SF fraction and the compressibility. Using these two observables the ($\mu/U$-$J Z/U$)-phase diagram is then determined, which displayed a round shape of the Mott lobes, a direct MI-SF transition for small disorder and an intervening BG phase at larger disorder. It should be noted that although the starting point of the calculation in \cite{Buon07,Buon09}, the LMF approximation, is identical to ours, the usage 
of a different criterion to identify the phases leads to a phase diagram that differs significantly from the one predicted by us.

A multi-site LMF theory is introduced in \cite{Pisa11}, where every plaque of two by two sites is treated quantum, which keeps the spatial correlation therein. Instead of single sites these plaques are coupled in a LMF way (analogous to section \ref{LMFtheory}). The Mott lobe is determined for both, the single-site and multi-site LMF theory, on the basis of the condensate fraction. In agreement with \cite{Buon07} and SMF theory it shows a round shape at the tip. The multi-site LMF theory predicts a larger MI region than the single-site LMF theory. Note that the condensate fraction smoothly approaches zero, analogous to our observations on the SF order parameter and the compressibility made in
section \ref{section:LSPC}; a linear fit is used to determine the transition point.

In \cite{Lin12} the so-called the Gutzwiller projected variational techniques is introduced in order to determine a canonical transformation of the quantum Hamiltonian, which requires the truncation of the hopping term. Thus, it is possible to minimize the expectation value of the transformed Hamiltonian in Gutzwiller type local mean field states with respect to its variational parameters. Finally, the SF stiffness and the compressibility yield the phase diagram, which shows a remarkably narrow BG region between the MI and the SF phase.

In all mean field calculations mentioned the tip of the Mott lobe is predicted for far higher values of the tunnelling rate as our results based on the LMF cluster analysis, and results of QMC or strong coupling expansion methods, as listed in Table \ref{quantumComparison}.

\subsection{The probability distribution of the local SF parameter}\label{section:PD}

In this section we discuss the probability distribution (PD) of the
local SF order parameter $P(\psi)$. In the ordered case $\Delta=0$,
depicted in the first row, all local SF parameter are identical since
all sites have the same on-site energy. The averaged order parameter
$\overline{\psi}$, depicted as a blue cross, is identical to each
local SF parameter $\psi_i$ and the variance of this value is
zero. Therefore, the PD $P\left(\psi\right)$ is a delta function at
the values of $\overline{\psi}$. Within the MI region the local SF
parameter is zero everywhere and thus the PD is a delta function at
$\psi=0$. In the SF regime still the PD is a delta function but at
positive $\psi$, which increases with the tunnelling strength.

\begin{figure}[!htb]
\begin{center}
\includegraphics[width=15cm]{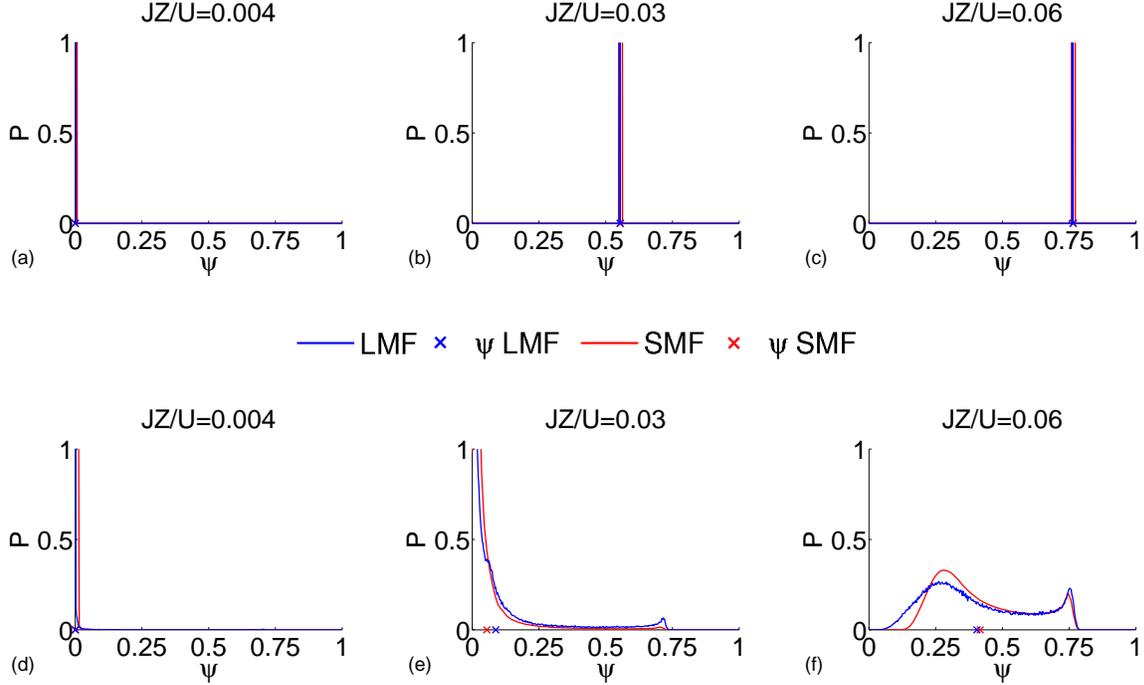}
\caption{\label{HistM} Probability distribution $P\left(\psi\right)$ 
of the local SF parameter at fixed chemical potential ($\mu/U=1.05$)
for different tunnelling rates $J$, (\textbf{top}) in the ordered case
($\Delta=0$) and (\textbf{bottom}) in the disordered case
($\Delta/U=0.6$) as predicted by the LMF (blue) and SMF (red) theory.
The crosses ($\times$) at the $\psi$-axis represent the mean of the
PD, which is the average SF order parameter $\overline{\psi}$. In the
ordered case the PD is a delta function. With disorder all local SF
parameters are zero for the MI; the PD in the BG phase is a
superposition of a delta function at $\psi=0$ and a SF tail; the SF
phase is characterized by a broad distribution of positive non-zero
local SF parameters.}
\end{center}
\end{figure}

This situation changes if disorder is introduced, since then the
on-site energy is different on every site resulting in a variety of
different values of the local SF parameter $\psi_i$. In the MI regime
the PD is a sharp delta function at $\psi=0$ still and becomes a broad
distribution in the BG and SF phase. In the BG phase sites with zero
local SF parameter (corresponding to MI sites with $S_i=0$) coexist
with sites, which have non-zero local SF parameter (corresponding to
SF sites with $S_i=1$) and where called SF islands before. In the SF
regime the PD is a broad distribution representing the variety of
positive values for the local SF parameter. Due to this characteristic
behaviour the PD can be written as a superposition of a delta function
at $\psi=0$ and a broad distribution representing the values $\psi>0$:
\begin{equation}\label{Psuperposition}
 P\left(\psi \right)=a \delta \left(\psi \right) + P_{\rm{SF}}\left(\psi \right).
\end{equation}
We denote this distribution $P_{\rm{SF}}\left(\psi
\right)$, since it represents sites, which we refereed to as SF sites
$(S_i=1)$ before:

Starting with the same Hamiltonian \eqref{HSMF} as LMF, SMF theory
introduces the additional approximation \eqref{SMFRestriction}
yielding in a self-consistent equation for the PD itself, which is
given by equation \eqref{SelfEqnP}. The resulting distributions are
shown in Figure \ref{HistM} in red. Whereas no deviations between LMF
and SMF theory occur in the ordered case ($\Delta=0$), they become
visible in the disordered case. For $\Delta>0$ oth distribution
have similar shapea, but especially for small values of
$\psi$, which are crucial for the determination of the phase
transition, they differ significantly in the BG and SF phase. 

In order to identify the reason for these deviations we scrutinized
the validity of the additional SMF restriction
\eqref{SMFRestriction}. We checked its validity by comparing the LMF
results for the product of the PD of two different sites
$P\left(\psi_i\right) P\left(\psi_j\right)$ with the pair PD $\mathcal
P_Z \left(\psi_i,\psi_j\right)$ and determined their deviation
$\Delta_P$ in \ref{section:assumptionSMF}. As shown in Figure
\ref{AbweichungP1P2} the approximation \eqref{SMFRestriction} is best
in the MI regime and the BG for very small tunnelling rates. But for
increasing $J Z/U$ is becomes worse and especially at the phase
boundary it fails. In Figure \ref{P1P2} where $P\left(\psi_i\right)
P\left(\psi_j\right)$ are $\mathcal P_Z
\left(\psi_i,\psi_j\right)$ are shown for different parameters it is
visible that they disagree especially for small values of the
$\psi$. This disagreement is due to the presence of correlations of the
local SF parameters $\psi_i$ at different sites in the vicinity of the
transition points, which are neglected in SMF theory.

\section{Conclusion}\label{section:Conclusion}

In this paper we have introduced a new criterion to identify the different phases of the disordered BHM in $d\ge2$ on the basis of the complete set of local boson occupation numbers $\{n_i\}$ of each sample and applied it to the ground states calculated using the LMF approximation. In the MI phase all $\langle\O[i]{n}\rangle$ are integer, in the BG phase some of them are non-integer and form SF clusters in a MI background and in the SF phase at least one of these clusters percolates. The emergence of SF clusters, with an average lateral size that is expected to be of the order of the SF correlation length, have a finite density, which gives rise to a non-vanishing mean of the average SF parameters although the system is not SF. The latter happens only when these SF clusters percolate, which is the hallmark of the BG-SF transition. Moreover, the SF clusters have a fluctuating boson occupation number resulting in a small but non-vanishing 
compressibility. Thus their appearance is the indicator of the MI-BG transition, i.e.\ from the incompressible ($\kappa=0$) to the compressible ($\kappa>0$) phase. Consequently, the BG phase displays arbitrarily small but non-vanishing values for $\overline{\psi}$ and $\kappa$ and all approaches to determine the LMF phase boundaries of the disordered system on the basis of the site and disorder averaged parameters, like the SF order parameter $\overline{\psi}$, the compressibility $\kappa$, the SF or condensate fraction, overestimates the SF and MI phases substantially and are doomed to fail: the putative phase boundaries move systematically and substantially when increasing or decreasing the threshold only by a small amount.

The resulting cluster analysis phase diagram for a fixed commensurate density $\overline{n}=1$ is in excellent agreement with the prediction of QMC simulations, not only qualitatively in reproducing the characteristic shape of the SF region in the ($\Delta$-$U$)-diagram, but also quantitatively within the numerical error bars. This is remarkably, since other LMF approaches using averaged quantities, like the mean SF order parameter or the compressibility as indicators predict much larger MI regions in the phase diagram or even fail to identify the BG transition, since the used indicator varies smoothly at the expected phase transition. Small deviations between QMC calculations and our LMF cluster analysis might be due to the fact that the local occupation numbers calculated by using the LMF approximation deviate in some regions of the phase diagram from the exact expectation values. Obviously it would be desirable to calculate the latter by QMC simulations and to perform the cluster analysis we propose to 
these data.

The questions that immediately arises in this context is: 1) Is the MI-BG transition in the disordered BHM exactly where SF sites occur? And more interestingly 2) Is the BG-SF transition actually identical with the percolation transition of SF clusters?

Concerning question 1): Although not proven rigorously, the MI-BG transition is supposed to occur, where the gap $E_{g/2}$, i.e. the energy for particle-hole excitations, of the pure, ordered BHM is equal to the disorder strength $\Delta$ \cite{Fish89}. It seems plausible that when this happens individual sites or small clusters will occur, where the addition or removal of a particle does not cost energy and thus the local boson occupation number fluctuates, i.e. $\langle\O[i]{n}\rangle$ is non-integer. This is how we propose to identify the MI-BG transition.

Concerning question 2) we argued in section \ref{section:PhasesLMF} in the basis of the BHM to quantum rotor models that the SF stiffness will always vanish as long as SF clusters with a MI background do not percolate. This BG situation then is reminiscent of a $d+1$-dimensional, classical XY model with columnar disorder, in a state with (quasi) phase ordered finite clusters in a phase disordered background. Application of a phase twist at the system boundaries will only cost a macroscopic amount of energy when the ordered regions actually percolate - in the SF region. Note that this argument is based upon the existence of true long-range order in the SF phase of the pure, ordered BHM, thus we expect it to be valid for $d\ge2$.

A complementary picture is based on the path-integral computation of the SF density \cite{Poll87}, which is used in QMC simulations to identify SF order. The SF density or stiffness is proportional to the mean-square of the winding number of boson world lines in the path integral representation.  When on average a finite fraction of boson world lines wrap around the whole system, the mean-square winding number is positive and the system is SF. To wrap around the whole system (with periodic boundary conditions), a boson world line, on its way through imaginary time, has to move along a path that traverses the whole system, thus attributing particle number fluctuations to the individual sites of this path. These sites will consequently attain non-integer expectation values for the boson occupation numbers $\langle\O[i]{n}\rangle$ (since for some time the boson was there and for some time not), thus in the end there must be at least one percolating SF cluster in the system.

It should be noted that other quantum phase transitions of disordered systems are naturally percolation transitions, too: The critical point
of the random transverse Ising model is governed by an infinite randomness fixed point (in $d\ge1$ dimensions \cite{Fish95,Fish99,Youn98,Kova11}), which signals the percolation of strongly coupled clusters that away from criticality constitute the Griffiths phase. The percolation transition that we observe in our calculations falls into the universality class of a conventional, 2d site percolation - which means it does not carry the signature of the critical properties of the proper BG-SF transition. This is most probably a consequence of the LMF approximation that we use, since it does not properly account for spatial correlations - if applied to the exact ground state one would expect the critical exponents of the percolation transition to be related to the critical exponents of the BG-SF transition.

In addition to providing an intuitive picture and a deeper understanding of the underlying physics of the phase transitions in the BHM the cluster analysis may serve as a reliable tool to locate the transitions in situations, in which the application of other criteria to discriminate the different phase might lead to erroneous predictions - as for instance in LMF theories. Finally, since experiments recently reached the regime of single site detection \cite{Bakr10, Sher10} and are now able to observe the particle numbers at each site, an experimental application of the cluster analysis that we propose appears in reach.
 
\newpage
\appendix

\section{Local SF order parameter correlations}\label{section:assumptionSMF}

\begin{figure}[!htb]
\begin{center}
 \includegraphics[width=10cm]{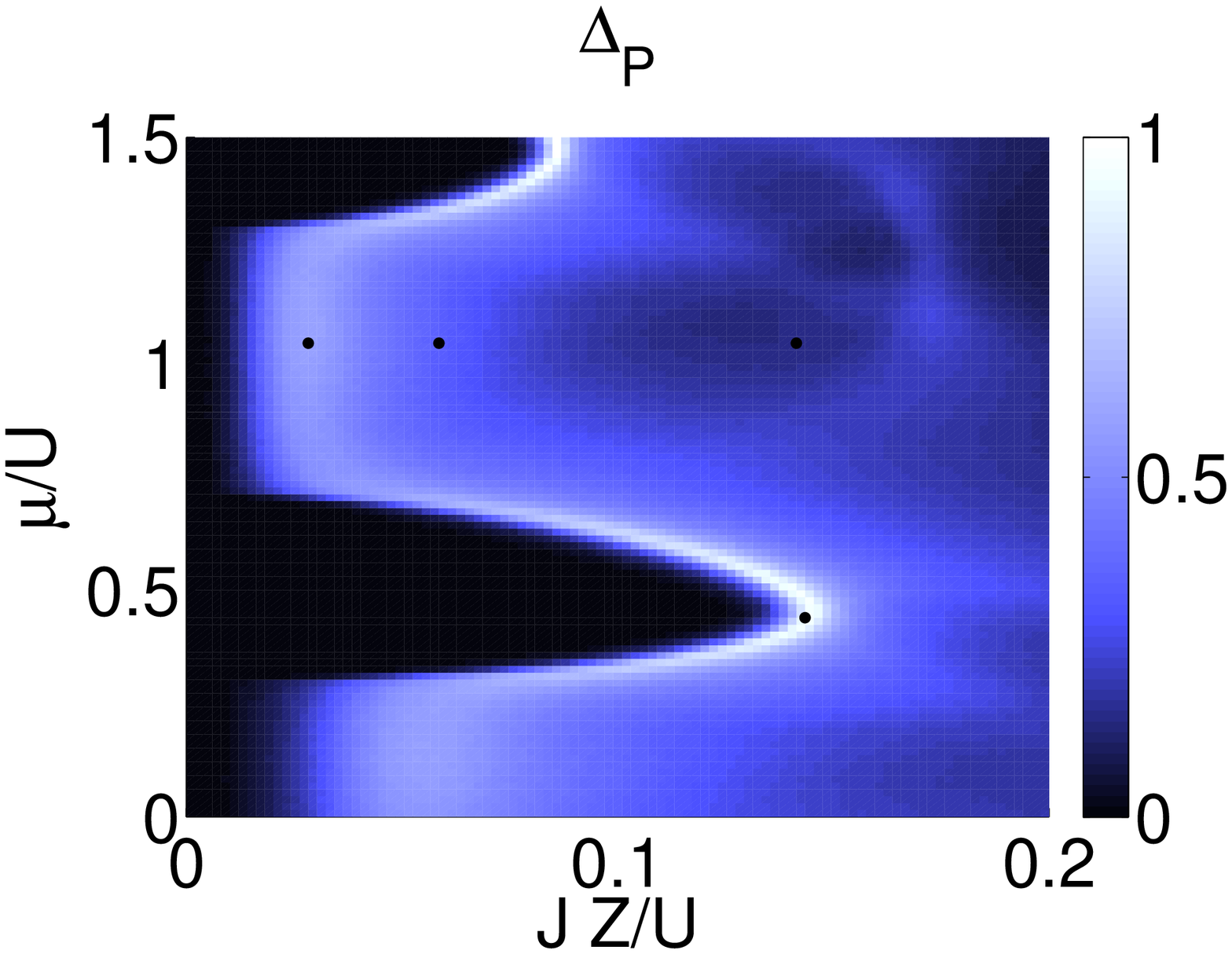}
 \caption{\label{AbweichungP1P2}The deviation $\Delta_P$ between $P\left(\psi_i\right) P\left(\psi_j\right)$ and $\mathcal P_Z \left(\psi_i,\psi_j\right)$ is shown in dependence of the system parameter. In the MI regime and for very small tunnelling rates the deviations are small, whereas they grow at the phase transitions and in the SF regime. The black dots make the parameters used in Figure \ref{P1P2} along a line $\mu/U=1.0455$ and at the tip of the Mott lobes, where the deviation is maximal.}
\end{center}
\end{figure}

\begin{figure}[!htb]
\begin{center}
\includegraphics[width=15cm]{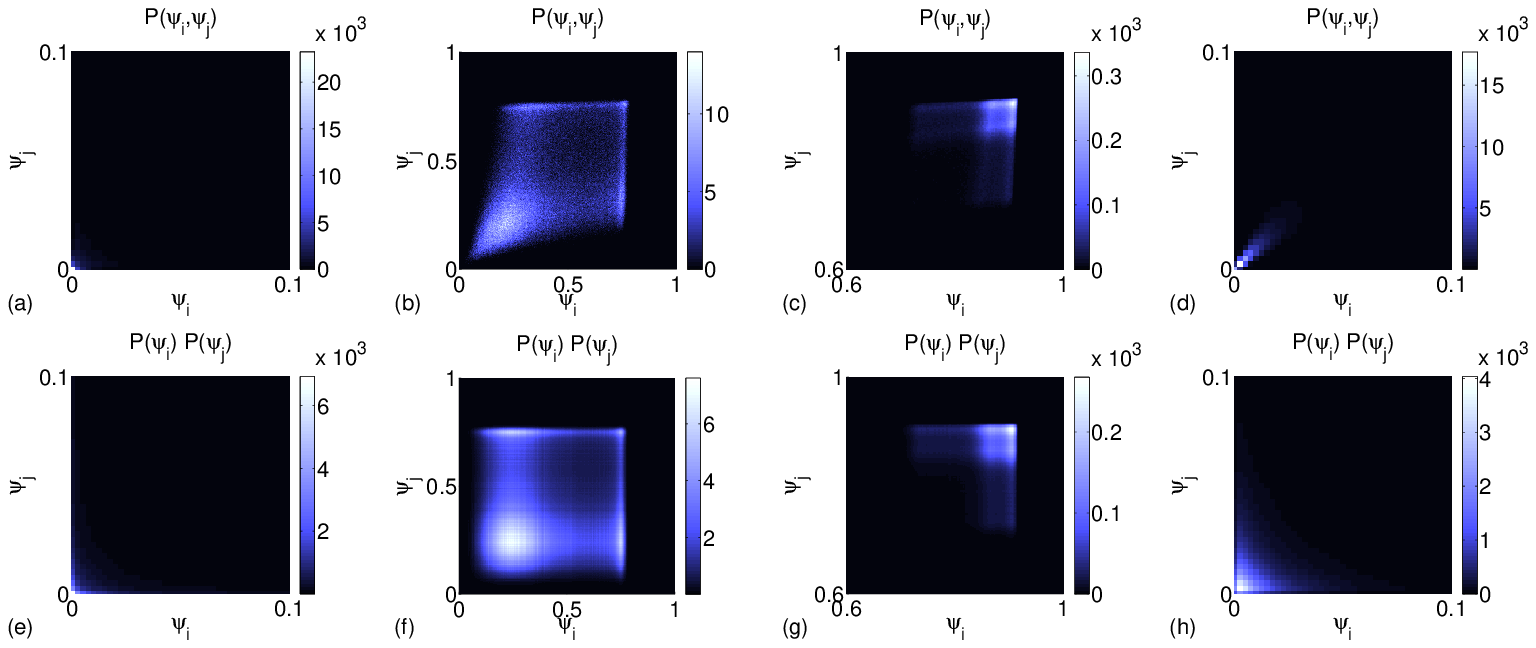}
\caption{\label{P1P2} The first column shows $\mathcal P_Z\left(\psi_i,\psi_j\right)$ and the second $P\left(\psi_i\right) P\left(\psi_j\right)$ in the disorder case ($\Delta/U=0.6$). In the first row the parameters are given by $JZ/U=0.0283,\;\mu/U=1.0455$ followed by $JZ/U= 0.0586,\; \mu/U=1.0455$ and $JZ/U=0.1414,\; \mu/U=1.0455$ and $JZ/U=0.1434,\; \mu/U=0.4394$ in the last row corresponding the black dots in Figure \ref{AbweichungP1P2}.}
\end{center}
\end{figure}

In this appendix we check the statistical independence assumption underlying SMF theory. Additional to the LMF approximation \eqref{LMF} made in the tunnelling part in the Hamiltonian, SMF theory assumes that the local SF parameters $\psi_1,\ldots,\psi_Z$ of the $Z$ neighbours of a chosen site $i$ are uncorrelated and identically distributed which is introduced by the approximation \eqref{SMFRestriction}. On the basis of LMF calculations we want to test this approximation by comparing $P\left(\psi_i\right) P\left(\psi_j\right)$ and $\mathcal P_Z\left(\psi_i,\psi_j\right)$, of which some examples are shown in
Figure \ref{P1P2}. The function $P\left(\psi_i\right)P\left(\psi_j\right)$ is the product of the PD, describing the distribution of the local SF parameter as discussed in section \ref{section:PD}. The function $\mathcal P_Z \left(\psi_i,\psi_j\right)$ is the PD of pairs $\left(\psi_i,\psi_j\right)$, where $i$ and $j$ are neighbouring sites. It represents the probability of having a specific value for the pair $\left(\psi_i,\psi_j\right)$. Exactly as $P\left(\psi_j\right)$ the $\mathcal P_Z \left(\psi_i,\psi_j\right)$ is computed for every realization of disorder and finally averaged. Both distributions
should coincide if the assumption \eqref{SMFRestriction} is valid. In Figure \ref{AbweichungP1P2} the integral difference
\begin{equation}
\Delta_P=\int \d\psi_i \int\d \psi_j |P\left(\psi_i\right) P\left(\psi_j\right)-\mathcal P_Z \left(\psi_i,\psi_j\right)| 
\end{equation}
of both distributions is shown in parameter space.

In the MI region, where the $P\left(\psi\right)$ is a delta function at $\psi=0$ and for very small tunnelling rate $J Z/U$ the deviations are small, whereas they are significant in the region of the phase transition and in the SF regime. For illustration both PDs are shown in Figure \ref{P1P2} along a line of $\mu/U=1.0455$ and at the tip of the Mott lobe, where the deviation $\Delta_P$ reaches its maximal value (corresponding to the black dots in Figure \ref{AbweichungP1P2}). Additional to the fact that all distributions are symmetric naturally, $P\left(\psi_i\right) P\left(\psi_j\right)$ shows a rectangular symmetry, which intrinsically follows from the fact that it is a product of the same PD $P\left(\psi\right)$. The PD $\mathcal P_Z \left(\psi_i \psi_j\right)$ containing further information of the occurring pairs shows systematic deviations. Whereas the values on the diagonal are reproduced quite well, the off-diagonal contributions are squeezed to the diagonal. This is especially pronounced in 
Figure \ref{P1P2} (d) and (h), which corresponds to the tip of the Mott lobe. These mean differences in comparison with $P\left(\psi_i\right) P\left(\psi_j\right)$ can be observed for all parameters shown in Figure \ref{P1P2} and mainly occur in the regime of small local SF parameters. \mbox{Figure \ref{AbweichungP1P2}} illustrates that the assumption \eqref{SMFRestriction} made in SMF theory, is well fulfilled in the MI regime but becomes worse in the region of the phase transition. Whether this theory predicts the phase transition reliably in this regime, is therefore to question.

The deviations occurring for small SF parameters in the limit of small $\psi$ as shown in Figure \ref{HistM} have also been discussed in \cite{Biss09}. In this work the authors concluded that LMF theory overestimates the phase coherence in the BG regime. But this is also true for SMF theory, since it is based on the same approximation of the tunnelling term of the Hamiltonian. In this paper we resolved this apparent problem by interpreting the SF-BG phase transition as a percolation transition (c.f. section \ref{section:Commensurate filling} and \ref{section:fixed Disorder}).

\section{The characteristic shapes of the PD}\label{section:phasePD}
In section \ref{section:PD} we discussed three different shapes of the PD in the disordered case, which are depicted at the bottom of Figure \ref{HistM}. In the MI phase the PD is given by a delta function at $\psi=0$, whereas in the BG and SF phase a broad distribution occurs, which means that $P\left(\psi\right)$ can be represented by a superposition of a delta function at $\psi=0$ and a continuous part $P_{\rm{SF}}\left(\psi \right)$ caused by SF sites with $\psi>0$ as defined in equation \eqref{Psuperposition}, and in the following denoted as SF distribution. Here, we will identify regions of the three different shapes of $P\left(\psi\right)$ in parameter space and discuss their connection to the phase transitions determined in section \ref{section:Commensurate filling} and \ref{section:fixed Disorder}.
\begin{figure}[!htb]
\begin{center}
 \includegraphics[width=15cm]{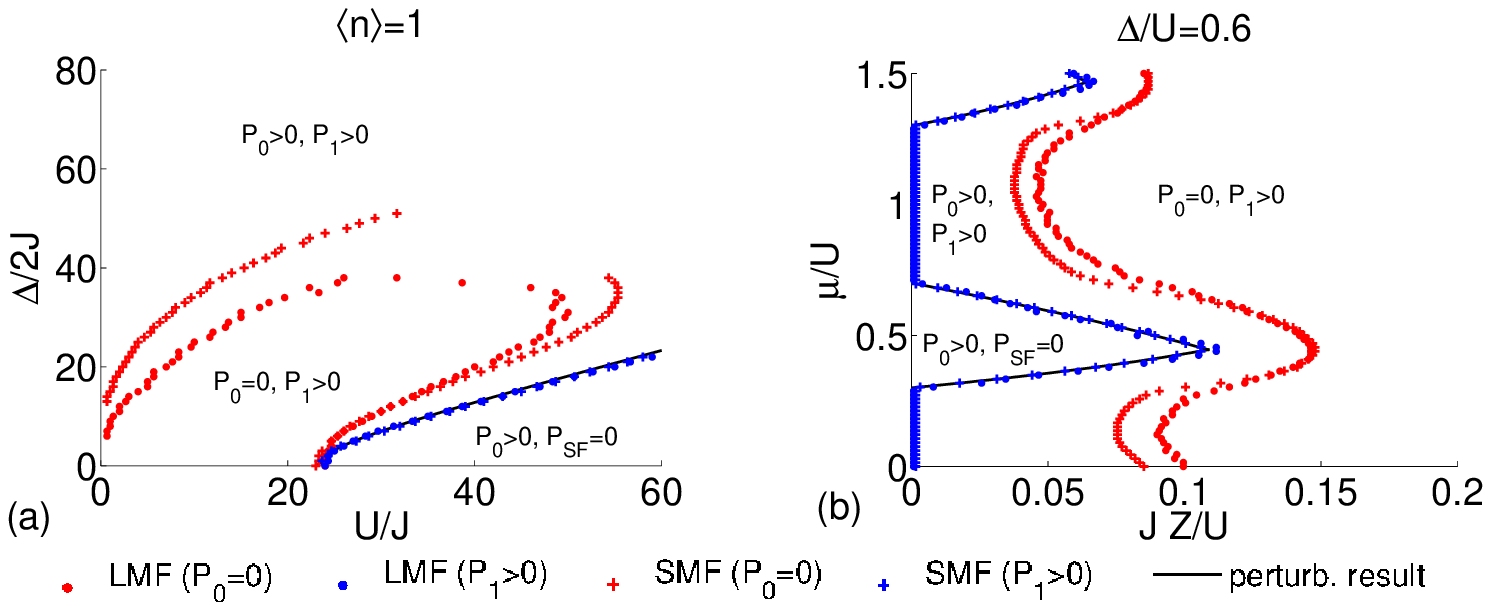}
 \caption{\label{PhasePe} Regions of the three characteristic shape of $P\left(\psi\right)$ for fixed density $\langle \O[]{n}\rangle=1$ {\bf (left)} and fixed disorder $\Delta/U=0.6$ {\bf (right)}. The black line represents the MI-BG transition according to the perturbative result given by \eqref{perturbative}. Inside of the blue line $P\left(\psi\right)$ consists only of a delta function at $\psi=0$. Within the region bounded by the red line on left and to the right of the red line on the right $P(\psi)$ only has a continuous part $P_{\rm{SF}}\left(\psi \right)$. In other parts of the parameter space is is a superposition of the  delta function at $\psi=0$ and the $P_{\rm{SF}}\left(\psi \right)$.}
\end{center}
\end{figure}

Numerically we identify two different benchmarks of the histograms representing $P\left(\psi\right)$ that we generate: The first one is the value of the histogram at the first bin, $P_0$, representing the potential delta peak of $P\left(\psi\right)$. The second characteristic point is the value of the histogram at the second bin, $P_1$, which is given by $P_1=P\left(\psi=\delta_{\psi}\right)$ with $\delta_{\psi}$ being the bin size of the histogram representing $P\left(\psi\right)$ ($\delta_{\psi}=0.0025$ in LMF and $\delta_{\psi}=0.015$ in SMF theory).

In Figure \ref{PhasePe} the regions determined on the basis of these benchmarks are shown for fixed density $\overline{n}=1$ on the left and fixed disorder strength $\Delta/U=0.6$ on the right. The LMF results are depicted with circles, SMF results with crosses. The blue curves enclose the regions in which $P\left(\psi\right)$ is just a delta function at $\psi=0$ (numerically: $P_0>0$ and $P_1<10^{-3}$), i.e. the system contains only MI sites and is therefore in the MI phase. The red curves delimit the regions, in which the delta function part of $P\left(\psi\right)$ vanishes (numerically $P_0<10^{-6}$), which means that all sites are SF sites. In the remaining part of the phase diagram $P\left(\psi\right)$ consists of a superposition of a delta function at $\psi=0$ and a continuous part $P_{\rm{SF}}\left(\psi \right)$, i.e. the system has MI and SF sites. In this region the system can either be in the BG phase, or, if the SF sites percolate, in the SF phase. Therefore only the MI-BG phase boundary can be 
extracted from the characteristics of the probability distribution of the local order parameter, $P\left(\psi\right)$, but not the BG-SF boundary. 

LMF and SMF theory coincide very well at the blue line, which describes the occurrence of the SF distribution. Deviations are visible at the red line, where the delta function at $\psi=0$ disappears and the PD is purely given by $P_{\rm{SF}}\left(\psi \right)$. While these discrepancies are rather small for small disorder strengths and $U/J>23$, they enlarge with increasing disorder. In \ref{section:assumptionSMF} we tested the validity the SMF approximation \eqref{SMFRestriction} by comparing the LMF results for the product of the PD of two different sites $P\left(\psi_i\right) P\left(\psi_j\right)$ with the pair PD $\mathcal P_Z \left(\psi_i,\psi_j\right)$. In Figure \ref{AbweichungP1P2} its deviation $\Delta_P$ is shown in dependence of $\mu/U$ and $J Z/U$ for fixed disorder $\Delta/U=0.6$. In comparison with the diagram on the right in Figure \ref{PhasePe} it is obvious that in the region of the blue line the SMF assumption \eqref{SMFRestriction} for $P\left(\psi\right)$ is valid. However, at the red curve 
deviations between LMF and SMF theory occur, since close to the BG-SF phase transition the approximation \eqref{SMFRestriction} is expected to be invalid.

\end{document}